\documentclass[ALICE,manyauthors]{cernphprep}
\usepackage[comma,square,numbers,sort&compress]{natbib}
\usepackage{hyperref}
\usepackage{lineno}
\usepackage{booktabs}
\usepackage{xspace}
\usepackage{multirow}
\usepackage[T1]{fontenc}
\usepackage{orcidlink}

\begin{document}
%

\newcommand{\pp}           {pp\xspace}
\newcommand{\ppbar}        {\mbox{$\mathrm {p\overline{p}}$}\xspace}
\newcommand{\XeXe}         {\mbox{Xe--Xe}\xspace}
\newcommand{\PbPb}         {\mbox{Pb--Pb}\xspace}
\newcommand{\pA}           {\mbox{pA}\xspace}
\newcommand{\pPb}          {\mbox{p--Pb}\xspace}
\newcommand{\AuAu}         {\mbox{Au--Au}\xspace}
\newcommand{\dAu}          {\mbox{d--Au}\xspace}

\newcommand{\s}            {\ensuremath{\sqrt{s}}\xspace}
\newcommand{\snn}          {\ensuremath{\sqrt{s_{\mathrm{NN}}}}\xspace}
\newcommand{\pt}           {\ensuremath{p_{\rm T}}\xspace}
\newcommand{\meanpt}       {$\langle p_{\mathrm{T}}\rangle$\xspace}
\newcommand{\ycms}         {\ensuremath{y_{\rm CMS}}\xspace}
\newcommand{\ylab}         {\ensuremath{y_{\rm lab}}\xspace}
\newcommand{\etarange}[1]  {\mbox{$\left | \eta \right |~<~#1$}}
\newcommand{\yrange}[1]    {\mbox{$\left | y \right |~<~#1$}}
\newcommand{\dndy}         {\ensuremath{\mathrm{d}N_\mathrm{ch}/\mathrm{d}y}\xspace}
\newcommand{\dndeta}       {\ensuremath{\mathrm{d}N_\mathrm{ch}/\mathrm{d}\eta}\xspace}
\newcommand{\avdndeta}     {\ensuremath{\langle\dndeta\rangle}\xspace}
\newcommand{\dNdy}         {\ensuremath{\mathrm{d}N_\mathrm{ch}/\mathrm{d}y}\xspace}
\newcommand{\Npart}        {\ensuremath{N_\mathrm{part}}\xspace}
\newcommand{\Ncoll}        {\ensuremath{N_\mathrm{coll}}\xspace}
\newcommand{\dEdx}         {\ensuremath{\textrm{d}E/\textrm{d}x}\xspace}
\newcommand{\RpPb}         {\ensuremath{R_{\rm pPb}}\xspace}

\newcommand{\nineH}        {$\sqrt{s}~=~0.9$~Te\kern-.1emV\xspace}
\newcommand{\seven}        {$\sqrt{s}~=~7$~Te\kern-.1emV\xspace}
\newcommand{\twoH}         {$\sqrt{s}~=~0.2$~Te\kern-.1emV\xspace}
\newcommand{\twosevensix}  {$\sqrt{s}~=~2.76$~Te\kern-.1emV\xspace}
\newcommand{\five}         {$\sqrt{s}~=~5.02$~Te\kern-.1emV\xspace}
\newcommand{\twosevensixnn}{$\sqrt{s_{\mathrm{NN}}}~=~2.76$~Te\kern-.1emV\xspace}
\newcommand{\fivenn}       {$\sqrt{s_{\mathrm{NN}}}~=~5.02$~Te\kern-.1emV\xspace}
\newcommand{\LT}           {L{\'e}vy-Tsallis\xspace}
\newcommand{\GeVc}         {Ge\kern-.1emV/$c$\xspace}
\newcommand{\GeVcc}        {Ge\kern-.2emV/$c^2$\xspace}
\newcommand{\MeVc}         {Me\kern-.1emV/$c$\xspace}
\newcommand{\MeVcc}        {Me\kern-.2emV/$c^2$\xspace}
\newcommand{\TeV}          {Te\kern-.1emV\xspace}
\newcommand{\GeV}          {Ge\kern-.1emV\xspace}
\newcommand{\MeV}          {Me\kern-.1emV\xspace}
\newcommand{\GeVmass}      {Ge\kern-.2emV/$c^2$\xspace}
\newcommand{\MeVmass}      {Me\kern-.2emV/$c^2$\xspace}
\newcommand{\lumi}         {\ensuremath{\mathcal{L}}\xspace}

\newcommand{\ITS}          {\rm{ITS}\xspace}
\newcommand{\TOF}          {\rm{TOF}\xspace}
\newcommand{\ZDC}          {\rm{ZDC}\xspace}
\newcommand{\ZDCs}         {\rm{ZDCs}\xspace}
\newcommand{\ZNA}          {\rm{ZNA}\xspace}
\newcommand{\ZNC}          {\rm{ZNC}\xspace}
\newcommand{\SPD}          {\rm{SPD}\xspace}
\newcommand{\SDD}          {\rm{SDD}\xspace}
\newcommand{\SSD}          {\rm{SSD}\xspace}
\newcommand{\TPC}          {\rm{TPC}\xspace}
\newcommand{\TRD}          {\rm{TRD}\xspace}
\newcommand{\VZERO}        {\rm{V0}\xspace}
\newcommand{\VZEROA}       {\rm{V0A}\xspace}
\newcommand{\VZEROC}       {\rm{V0C}\xspace}
\newcommand{\Vdecay} 	   {\ensuremath{V^{0}}\xspace}

\newcommand{\ee}           {\ensuremath{\mathrm{e^{+}e^{-}}}} 
\newcommand{\pip}          {\ensuremath{\pi^{+}}\xspace}
\newcommand{\pim}          {\ensuremath{\pi^{-}}\xspace}
\newcommand{\kap}          {\ensuremath{\rm{K}^{+}}\xspace}
\newcommand{\kam}          {\ensuremath{\rm{K}^{-}}\xspace}
\newcommand{\pbar}         {\ensuremath{\rm\overline{p}}\xspace}
\newcommand{\kzero}        {\ensuremath{{\rm K}^{0}_{\rm{S}}}\xspace}
\newcommand{\lmb}          {\ensuremath{\Lambda}\xspace}
\newcommand{\almb}         {\ensuremath{\overline{\Lambda}}\xspace}
\newcommand{\Om}           {\ensuremath{\Omega^-}\xspace}
\newcommand{\Mo}           {\ensuremath{\overline{\Omega}^+}\xspace}
\newcommand{\X}            {\ensuremath{\Xi^-}\xspace}
\newcommand{\Ix}           {\ensuremath{\overline{\Xi}^+}\xspace}
\newcommand{\Xis}          {\ensuremath{\Xi^{\pm}}\xspace}
\newcommand{\Oms}          {\ensuremath{\Omega^{\pm}}\xspace}
\newcommand{\degree}       {\ensuremath{^{\rm o}}\xspace}

\newcommand{\mee}{\ensuremath{m_{\mathrm{ee}}}\xspace}

\newcommand{\ptee}{\ensuremath{p_{\mathrm{T,ee}}}\xspace}
\newcommand{\sNN}{\ensuremath{\sqrt{s_{\mathrm{NN}}}}\xspace}
\newcommand{\DCAee}{\ensuremath{\mathrm{DCA_{ee}}}\xspace}

\newcommand{\Jpsi}{\ensuremath{\mathrm{J}/\psi }\xspace}

\newcommand{\qqbar}{\ensuremath{\mathrm {q\bar{q}}}\xspace}
\newcommand{\bbbar}{\ensuremath{\mathrm {b\bar{b}}}\xspace}
\newcommand{\ccbar}{\ensuremath{\mathrm {c\overline{c}}}\xspace}
\newcommand{\QQbar}{\ensuremath{\mathrm {Q\overline{Q}}}\xspace}
\newcommand{\DDbar}{\ensuremath{\mathrm {D\overline{D}}}\xspace}
\newcommand{\BBbar}{\ensuremath{\mathrm {B\overline{B}}}\xspace}

\newcommand{\cHc}{\ensuremath{\mathrm{c \rightarrow H_{c}}}\xspace}
\newcommand{\Hce}{\ensuremath{\mathrm{H_{c} \rightarrow e}}\xspace}

\begin{titlepage}
\PHyear{2024}       
\PHnumber{247}      
\PHdate{18 September}  

\title{Addendum: Dielectron production in proton--proton and proton--lead collisions at $\mathbf{\sqrt{\textit{s}_{\textup{NN}}} = 5.02}$~TeV}
\ShortTitle{Addendum: Dielectron production in pp and p--Pb at $\sqrt{s_{\rm NN}}=$ 5.02 TeV}   

\Collaboration{ALICE Collaboration\thanks{See Appendix~\ref{app:collab} for the list of collaboration members}}
\ShortAuthor{ALICE Collaboration} 

\begin{abstract}
This is an addendum to the article "Dielectron production in proton--proton and proton--lead collisions at $\sqrt{s_{\mathrm{NN}}}$ = 5.02 TeV" published in Phys. Rev. C \textbf{102} no. 5, (2020) 055204~\cite{ALICE:2020mfy}. We update the extracted charm cross section at midrapidity given in Table 3 and Figure 4 (left) of the original publication with the fragmentation fractions of charm quarks in pp collisions published in Ref.~\cite{ALICE:2021dhb}.
\end{abstract}
\end{titlepage}

\setcounter{page}{2} 


In Ref.~\cite{ALICE:2020mfy} the production cross section of charm and anti-charm (\ccbar) quark pairs was measured at midrapidity using a template fit to the measured dielectron yield in the invariant mass (\mee) region $1.1~<~\mee~<~2.7$~\GeVcc, the so-called intermediate mass range (IMR). The fit was performed simultaneously in the \mee and dielectron transverse momentum ($\ptee < 8$~\GeVc) dimensions, using either POWHEG~\cite{powheg1,powheg2,powheg3,powheg4} or PYTHIA~6~\cite{pythia64} as event generators for the templates of the charm and beauty components. Both generators give different shapes of the dielectron spectrum as a function of \mee and \ptee, however, they describe the data well after the fit.
To further extract the \ccbar and \bbbar cross sections at midrapidity, an extrapolation from the IMR to full phase space was performed using the functional shape given by the generators. To obtain the cross section, the probability of a charm or beauty quark fragmenting into hadrons, and those decaying with an electron in the final state was used to calculate the heavy-quark (\QQbar) cross sections from the heavy-flavour dielectron cross section. These so-called effective branching ratios BR($\mathrm{Q \rightarrow e}$) are a convolution of the fragmentation fractions (FF) of the heavy quarks and the decay branching ratios of open heavy-flavour hadrons into electrons. In the case of beauty, the effective branching ratios implemented in PYTHIA~6 were used~\cite{ALICE:2018fvj}. For charm, the central value of $\Gamma(\mathrm{c\rightarrow l + X}) / \Gamma(\mathrm{c\rightarrow X})$ from the 2020 edition of the PDG~\cite{ParticleDataGroup:2020ssz} was used, i.e. 9.6\%. 

With the measured FF in pp collisions at $\sqrt{s} = 5.02$~TeV at the LHC~\cite{ALICE:2021dhb}, the effective branching ratio of charm quarks into electrons can be reevaluated. This first measurement at LHC energies also extends the previous measurements by including the FF of the $\mathrm{\Xi_{c}}^{0}$. For the fragmentation fraction of charm quarks into $\Xi^{+}_{\mathrm{c}}$ baryons, $f(\mathrm{c} \rightarrow \Xi^{+}_{\mathrm{c}})$, the same value as for \mbox{$f(\mathrm{c} \rightarrow \Xi^{0}_{\mathrm c})$} is used, assuming isospin symmetry. This is confirmed by measurements in pp collisions at $\s = 13$~TeV~\cite{ALICE:2023sgl}. The values used are summarised in Table~\ref{tab:summary}. Listed are the fragmentation fractions of c quarks into charm hadrons ($\mathrm{H_{c}}$)~\cite{ALICE:2021dhb}, together with their branching ratios into electrons based on Ref.~\cite{ParticleDataGroup:2024cfk}. In the last column, the product of both is calculated, giving the probability of $\mathrm{c \rightarrow H_{c} \rightarrow e}$ for each charm hadron species. The absolute probability for a c quark to produce an electron after fragmentation and decay is then given by the sum over these values and amounts to $7.33^{+0.61}_{-0.66}$\%. This does not include the contribution of $\rm \Omega_{c}^{0}$, leading to an additional uncertainty of $-0.52\%$. This is added in quadrature to the lower uncertainty resulting in \mbox{$\mathrm{BR(c \to e)} = 7.33^{+0.61}_{-0.84}$},
which corresponds to $0.54^{+0.09}_{-0.12}$\% for the $\ccbar \rightarrow \ee$ case.

\begin{table}[ht!]
    \centering
    \caption{The charm quark fragmentation fractions into charm hadrons $f$(\cHc) from Ref.~\cite{ALICE:2021dhb} are listed together with the respective branching ratios of the hadrons into electron BR(\Hce) from Ref.~\cite{ParticleDataGroup:2024cfk}, as well as their product. The systematic and statistical uncertainties of $f$(\cHc) are added in quadrature, assuming they are uncorrelated.
    If available, the inclusive values for of the branching ratios with electrons in the final state are used. In the case of $\Xi_{\rm{c}}^{+}$ and $\Xi_{\rm{c}}^{0}$ only one decay with an electron in the final state is measured. In these cases, the BR for the respective channel is used.}
    \begin{tabular}{c c c c}
    \toprule
    $\mathrm{H_{c}}$          &  $f\mathrm{(\cHc)}$[\%] & $\mathrm{BR(\Hce)}$[\%] & $f\times\mathrm{BR}$[\%] \\
    \midrule
    $D^{0}$                   & $\mathrm{39.1 \pm 1.7 {\rm{(stat)}} ^{+2.5} _{-2.3} {\rm{(syst)}}}$             & $6.49\pm0.11$       & $2.54^{+ 0.20 }_{- 0.27 }$\\
    $D^{+}$                   & $\mathrm{17.3 \pm 1.8 {\rm{(stat)}} ^{+1.7} _{-1.7} {\rm{(syst)}}}$             & $16.07\pm0.3$       & $2.78^{+ 0.40 }_{- 0.45 }$\\
    $D^{+}_{\mathrm s}$       & $\mathrm{7.3  \pm 1.0 {\rm{(stat)}} ^{+1.9} _{-0.1} {\rm{(syst)}}}$             & $6.33\pm0.15$       & $0.46^{+ 0.14 }_{- 0.09 }$\\
    $\Lambda^{+}_{\mathrm c}$ & $\mathrm{20.4 \pm 1.3 {\rm{(stat)}} ^{+1.6} _{-1.6} {\rm{(syst)}}}$             & $4.06\pm0.13$       & $0.83^{+ 0.09 }_{- 0.09 }$\\
    $\Xi^{0}_{\mathrm c}$     & $\mathrm{8.0  \pm 1.2 {\rm{(stat)}} ^{+2.5} _{-2.3} {\rm{(syst)}}}$             & $1.94\pm0.55$  & $0.16^{+ 0.05 }_{- 0.05}$\\
    $\Xi^{+}_{\mathrm c}$     & $\mathrm{8.0  \pm 1.2 {\rm{(stat)}} ^{+2.5} _{-2.3} {\rm{(syst)}}}$             & $7\pm4$             & $0.56^{+ 0.37 }_{- 0.37 }$\\
    \cmidrule{4-4}
                              &                                               &                     & $7.33^{+0.61}_{-0.66}$ \\
    \bottomrule 
    &
    \end{tabular}
    \label{tab:summary}
\end{table}
The comparison with the previously used value of $\mathrm{BR(c \to e)}=9.6\%$ leads to a larger \ccbar production cross section at midrapidity ($\mathrm{d}\sigma_{\ccbar}/\mathrm{d}y|_{y=0}$) by a factor 1.72 as the change in $\mathrm{BR(c \to e)}$ enters the final results squared. The previously assigned uncertainty on BR($\ccbar\rightarrow\ee$) was 22\%, which is now reevaluated  to $^{+23}_{-17}$\%. 
The \ccbar cross sections in pp collisions at $\sqrt{s} = 7$ and 13 TeV~\cite{ALICE:2018fvj, ALICE:2018gev} can be updated with the same procedure.
The reevaluated \ccbar production cross sections at midrapidity are summarised in Table~\ref{tab:my_label}. Note that with respect to Refs.~\cite{ALICE:2018fvj, ALICE:2018gev, ALICE:2020mfy} only the \ccbar cross sections change.

\begin{table}[h]
    \centering
    \caption{\ccbar production cross sections at midrapidity extracted at different collision energies using the event generators PYTHIA~6 and POWHEG with an effective branching ratio of charm quarks to electrons based on fragmentation fractions measured by the ALICE collaboration~\cite{ALICE:2021dhb}.}
    \begin{tabular}{c c c}
    \toprule
    $\sqrt{s}$ (TeV) & \multicolumn{2}{c}{ $\sigma_{\mathrm {c\bar{c}}}|_{y=0}$ ($\mu$b)}  \\
    \midrule
    & {PYTHIA} & {POWHEG}  \\
    5.02 & $\rm 900 \pm 105(stat) \pm 45(syst)$ $^{+208}_{-152}$ (BR) & $\rm 1299 \pm 137(stat) \pm 65(syst)$ $^{+300}_{-220}$ (BR) \\
    7 & $\rm 1468 \pm 211(stat) \pm 249(syst)$ $^{+339}_{-248}$ (BR)  & $\rm 2150 \pm 266(stat) \pm 366(syst)$ $^{+497}_{-363}$ (BR)  \\
    13 & $\rm 1674 \pm 237(stat) \pm 241(syst)$ $^{+387}_{-283}$ (BR) & $\rm 2435 \pm 316(stat) \pm 351(syst)$ $^{+562}_{-411}$ (BR)  \\
 \bottomrule
  &
    \end{tabular}
    \label{tab:my_label}
\end{table}

\begin{figure}[ht]
    \centering
    \includegraphics[scale=0.7]{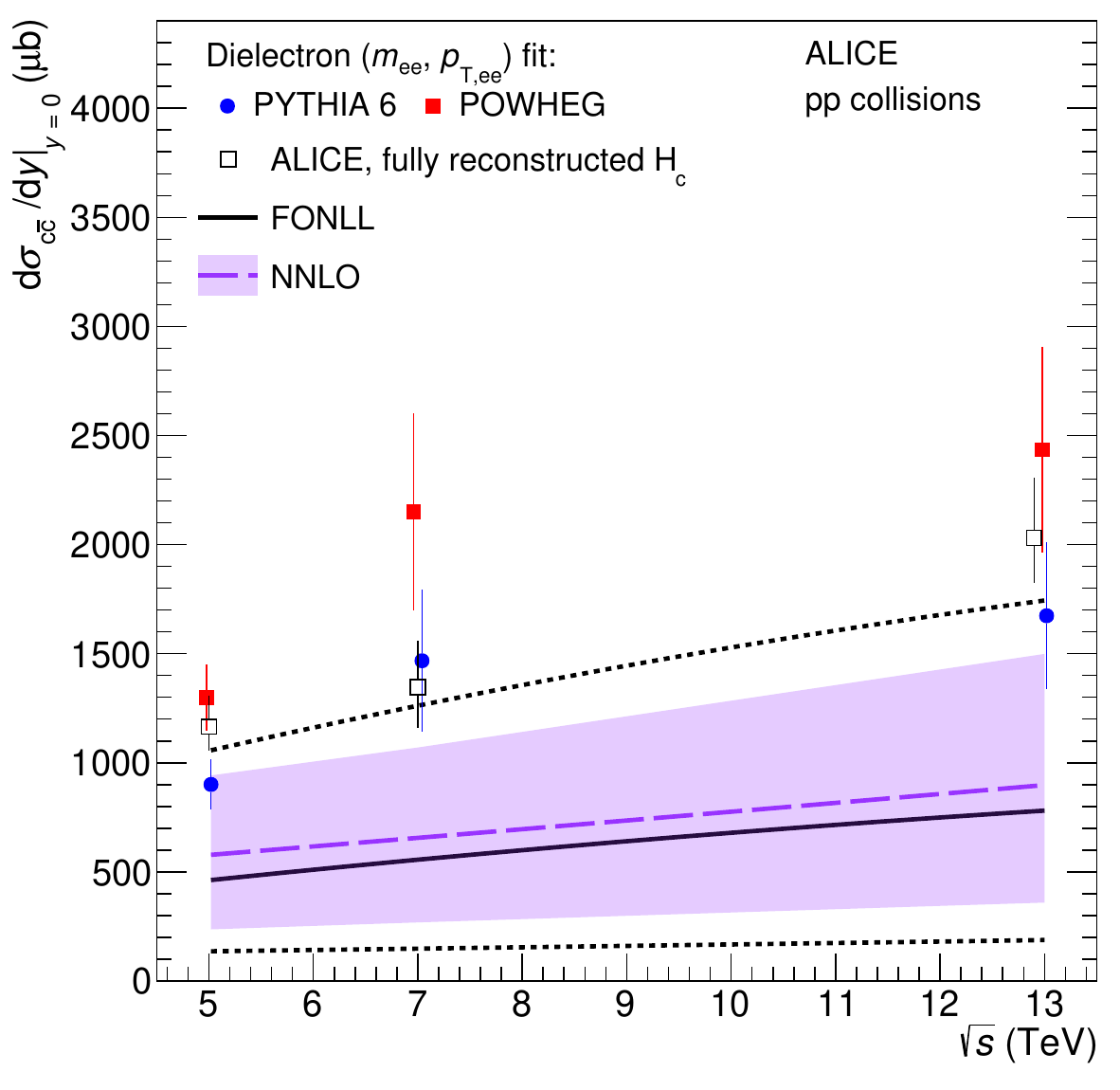}
    \caption{(Colour online) Cross sections at midrapidity for \ccbar production as a function of $\sqrt{s}$ in pp~collisions. The coloured markers represent the measured midrapidity cross sections at $\sqrt{s}$ = 5.02, 7, and 13 TeV which are derived using either PYTHIA 6 (blue circles) or POWHEG (red squares) simulations~\cite{ALICE:2018fvj, ALICE:2018gev, ALICE:2020mfy}. The systematic and statistical uncertainty of the data points are summed in quadrature and represented by vertical bars.
    The measurements in the dielectron analysis are compared to measurements performed with fully reconstructed charm hadrons ($\mathrm{H_{c}}$) at the same energies (open markers)~\cite{ALICE:2021dhb, ALICE:2023sgl}.
    In addition, measurements are compared to FONLL calculations (black solid line)~\cite{Cacciari:1998it, Cacciari:2001td, Cacciari:2012ny, Cacciari:2015fta} with corresponding model uncertainties (dashed lines), with next-to-next-to-leading-order (NNLO) calculations (magenta)~\cite{dEnterria:2016ids, dEnterria:2016yhy}.}
    \label{fig:csSummary}
\end{figure}

The compilation of the measured \ccbar cross sections as a function of $\sqrt{s}$ is shown in Fig.~\ref{fig:csSummary}, together with the complementary measurement of the charm production cross sections at midrapidity in the fully reconstructed charm hadron channels in pp collisions at $\sqrt{s} = 5.02$ and 7 TeV~\cite{ALICE:2021dhb}, as well as 13 TeV~\cite{ALICE:2023sgl}. Within uncertainties the dielectron measurements are consistent with the results using charm hadrons. The measurements are compared to predictions calculated in the fixed-order next-to-leading-log (FONLL) approach~\cite{Cacciari:1998it, Cacciari:2001td, Cacciari:2012ny, Cacciari:2015fta}. Shifting up the data points coherently, the $\sqrt{s}$ dependence of the cross section is unchanged with respect to the original publication and described by the FONLL calculations. While consistent with FONLL predictions within uncertainties, the measurements tend to be on the upper edge of the theoretical uncertainty band. Calculations at next-to-next-to-leading-order (NNLO) predict a consistently higher \ccbar cross section as a function of $\sqrt{s}$~\cite{dEnterria:2016ids, dEnterria:2016yhy}. The smaller uncertainties however increase the tension with the measurements.

\newpage

As for the beauty sector, enhanced baryon-to-meson ratios for beauty hadrons~\cite{LHCb:2019fns} and non-prompt charm hadrons~\cite{ALICE:2023wbx} have been measured in hadronic collisions at the LHC compared to results in e$^{+}$e$^{-}$ collisions at lower energies. Since the semi-leptonic branching ratios of the different beauty hadrons are very close to each other, these observations affect only the dielectron pairs involving at least one charm hadron decay ($\mathrm{b \to c \to e}$). Since no decays were forced in the creation of the templates, the fragmentation and decay in the relevant PYTHIA version are responsible for reproducing the measurements. A reweighting was performed to study the validity of the utilized templates to extract the cross sections. It was found that the overall effect on the shape of the beauty template is negligible in the phase space where the measurement is sensitive to the beauty production.


\newenvironment{acknowledgement}{\relax}{\relax}
\begin{acknowledgement}
\section*{Acknowledgements}

The ALICE Collaboration would like to thank all its engineers and technicians for their invaluable contributions to the construction of the experiment and the CERN accelerator teams for the outstanding performance of the LHC complex.
The ALICE Collaboration gratefully acknowledges the resources and support provided by all Grid centres and the Worldwide LHC Computing Grid (WLCG) collaboration.
The ALICE Collaboration acknowledges the following funding agencies for their support in building and running the ALICE detector:
A. I. Alikhanyan National Science Laboratory (Yerevan Physics Institute) Foundation (ANSL), State Committee of Science and World Federation of Scientists (WFS), Armenia;
Austrian Academy of Sciences, Austrian Science Fund (FWF): [M 2467-N36] and Nationalstiftung f\"{u}r Forschung, Technologie und Entwicklung, Austria;
Ministry of Communications and High Technologies, National Nuclear Research Center, Azerbaijan;
Conselho Nacional de Desenvolvimento Cient\'{\i}fico e Tecnol\'{o}gico (CNPq), Financiadora de Estudos e Projetos (Finep), Funda\c{c}\~{a}o de Amparo \`{a} Pesquisa do Estado de S\~{a}o Paulo (FAPESP) and Universidade Federal do Rio Grande do Sul (UFRGS), Brazil;
Bulgarian Ministry of Education and Science, within the National Roadmap for Research Infrastructures 2020-2027 (object CERN), Bulgaria;
Ministry of Education of China (MOEC) , Ministry of Science \& Technology of China (MSTC) and National Natural Science Foundation of China (NSFC), China;
Ministry of Science and Education and Croatian Science Foundation, Croatia;
Centro de Aplicaciones Tecnol\'{o}gicas y Desarrollo Nuclear (CEADEN), Cubaenerg\'{\i}a, Cuba;
Ministry of Education, Youth and Sports of the Czech Republic, Czech Republic;
The Danish Council for Independent Research | Natural Sciences, the VILLUM FONDEN and Danish National Research Foundation (DNRF), Denmark;
Helsinki Institute of Physics (HIP), Finland;
Commissariat \`{a} l'Energie Atomique (CEA) and Institut National de Physique Nucl\'{e}aire et de Physique des Particules (IN2P3) and Centre National de la Recherche Scientifique (CNRS), France;
Bundesministerium f\"{u}r Bildung und Forschung (BMBF) and GSI Helmholtzzentrum f\"{u}r Schwerionenforschung GmbH, Germany;
General Secretariat for Research and Technology, Ministry of Education, Research and Religions, Greece;
National Research, Development and Innovation Office, Hungary;
Department of Atomic Energy Government of India (DAE), Department of Science and Technology, Government of India (DST), University Grants Commission, Government of India (UGC) and Council of Scientific and Industrial Research (CSIR), India;
National Research and Innovation Agency - BRIN, Indonesia;
Istituto Nazionale di Fisica Nucleare (INFN), Italy;
Japanese Ministry of Education, Culture, Sports, Science and Technology (MEXT) and Japan Society for the Promotion of Science (JSPS) KAKENHI, Japan;
Consejo Nacional de Ciencia (CONACYT) y Tecnolog\'{i}a, through Fondo de Cooperaci\'{o}n Internacional en Ciencia y Tecnolog\'{i}a (FONCICYT) and Direcci\'{o}n General de Asuntos del Personal Academico (DGAPA), Mexico;
Nederlandse Organisatie voor Wetenschappelijk Onderzoek (NWO), Netherlands;
The Research Council of Norway, Norway;
Pontificia Universidad Cat\'{o}lica del Per\'{u}, Peru;
Ministry of Science and Higher Education, National Science Centre and WUT ID-UB, Poland;
Korea Institute of Science and Technology Information and National Research Foundation of Korea (NRF), Republic of Korea;
Ministry of Education and Scientific Research, Institute of Atomic Physics, Ministry of Research and Innovation and Institute of Atomic Physics and Universitatea Nationala de Stiinta si Tehnologie Politehnica Bucuresti, Romania;
Ministry of Education, Science, Research and Sport of the Slovak Republic, Slovakia;
National Research Foundation of South Africa, South Africa;
Swedish Research Council (VR) and Knut \& Alice Wallenberg Foundation (KAW), Sweden;
European Organization for Nuclear Research, Switzerland;
Suranaree University of Technology (SUT), National Science and Technology Development Agency (NSTDA) and National Science, Research and Innovation Fund (NSRF via PMU-B B05F650021), Thailand;
Turkish Energy, Nuclear and Mineral Research Agency (TENMAK), Turkey;
National Academy of  Sciences of Ukraine, Ukraine;
Science and Technology Facilities Council (STFC), United Kingdom;
National Science Foundation of the United States of America (NSF) and United States Department of Energy, Office of Nuclear Physics (DOE NP), United States of America.
In addition, individual groups or members have received support from:
Czech Science Foundation (grant no. 23-07499S), Czech Republic;
FORTE project, reg.\ no.\ CZ.02.01.01/00/22\_008/0004632, Czech Republic, co-funded by the European Union, Czech Republic;
European Research Council (grant no. 950692), European Union;
ICSC - Centro Nazionale di Ricerca in High Performance Computing, Big Data and Quantum Computing, European Union - NextGenerationEU;
Academy of Finland (Center of Excellence in Quark Matter) (grant nos. 346327, 346328), Finland.

\end{acknowledgement}

\bibliographystyle{utphys}   
\bibliography{bibliography}

\newpage
\appendix

%
%

\section{The ALICE Collaboration}
\label{app:collab}
\begin{flushleft} 
\small

S.~Acharya\,\orcidlink{0000-0002-9213-5329}\,$^{\rm 127}$, 
D.~Adamov\'{a}\,\orcidlink{0000-0002-0504-7428}\,$^{\rm 86}$, 
A.~Agarwal$^{\rm 135}$, 
G.~Aglieri Rinella\,\orcidlink{0000-0002-9611-3696}\,$^{\rm 32}$, 
L.~Aglietta\,\orcidlink{0009-0003-0763-6802}\,$^{\rm 24}$, 
M.~Agnello\,\orcidlink{0000-0002-0760-5075}\,$^{\rm 29}$, 
N.~Agrawal\,\orcidlink{0000-0003-0348-9836}\,$^{\rm 25}$, 
Z.~Ahammed\,\orcidlink{0000-0001-5241-7412}\,$^{\rm 135}$, 
S.~Ahmad\,\orcidlink{0000-0003-0497-5705}\,$^{\rm 15}$, 
S.U.~Ahn\,\orcidlink{0000-0001-8847-489X}\,$^{\rm 71}$
% ...

\section*{Affiliation Notes}

$^{\rm I}$ Deceased\\
$^{\rm II}$ Also at: Max-Planck-Institut fur Physik, Munich, Germany\\
$^{\rm III}$ Also at: Italian National Agency for New Technologies, Energy and Sustainable Economic Development (ENEA), Bologna, Italy\\
$^{\rm IV}$ Also at: Dipartimento DET del Politecnico di Torino, Turin, Italy\\
$^{\rm V}$ Also at: Yildiz Technical University, Istanbul, T\"{u}rkiye\\
$^{\rm VI}$ Also at: Department of Applied Physics, Aligarh Muslim University, Aligarh, India\\
$^{\rm VII}$ Also at: Institute of Theoretical Physics, University of Wroclaw, Poland\\
$^{\rm VIII}$ Also at: An institution covered by a cooperation agreement with CERN\\

\section*{Collaboration Institutes}

$^{1}$ A.I. Alikhanyan National Science Laboratory (Yerevan Physics Institute) Foundation, Yerevan, Armenia\\
$^{2}$ AGH University of Krakow, Cracow, Poland\\
$^{3}$ Bogolyubov Institute for Theoretical Physics, National Academy of Sciences of Ukraine, Kiev, Ukraine\\
$^{4}$ Bose Institute, Department of Physics  and Centre for Astroparticle Physics and Space Science (CAPSS), Kolkata, India\\
$^{5}$ California Polytechnic State University, San Luis Obispo, California, United States\\
$^{6}$ Central China Normal University, Wuhan, China\\
$^{7}$ Centro de Aplicaciones Tecnol\'{o}gicas y Desarrollo Nuclear (CEADEN), Havana, Cuba\\
$^{8}$ Centro de Investigaci\'{o}n y de Estudios Avanzados (CINVESTAV), Mexico City and M\'{e}rida, Mexico\\
$^{9}$ Chicago State University, Chicago, Illinois, United States\\
$^{10}$ China Institute of Atomic Energy, Beijing, China\\
$^{11}$ China University of Geosciences, Wuhan, China\\
$^{12}$ Chungbuk National University, Cheongju, Republic of Korea\\
$^{13}$ Comenius University Bratislava, Faculty of Mathematics, Physics and Informatics, Bratislava, Slovak Republic\\
$^{14}$ Creighton University, Omaha, Nebraska, United States\\
$^{15}$ Department of Physics, Aligarh Muslim University, Aligarh, India\\
$^{16}$ Department of Physics, Pusan National University, Pusan, Republic of Korea\\
$^{17}$ Department of Physics, Sejong University, Seoul, Republic of Korea\\
$^{18}$ Department of Physics, University of California, Berkeley, California, United States\\
$^{19}$ Department of Physics, University of Oslo, Oslo, Norway\\
$^{20}$ Department of Physics and Technology, University of Bergen, Bergen, Norway\\
$^{21}$ Dipartimento di Fisica, Universit\`{a} di Pavia, Pavia, Italy\\
$^{22}$ Dipartimento di Fisica dell'Universit\`{a} and Sezione INFN, Cagliari, Italy\\
$^{23}$ Dipartimento di Fisica dell'Universit\`{a} and Sezione INFN, Trieste, Italy\\
$^{24}$ Dipartimento di Fisica dell'Universit\`{a} and Sezione INFN, Turin, Italy\\
$^{25}$ Dipartimento di Fisica e Astronomia dell'Universit\`{a} and Sezione INFN, Bologna, Italy\\
$^{26}$ Dipartimento di Fisica e Astronomia dell'Universit\`{a} and Sezione INFN, Catania, Italy\\
$^{27}$ Dipartimento di Fisica e Astronomia dell'Universit\`{a} and Sezione INFN, Padova, Italy\\
$^{28}$ Dipartimento di Fisica `E.R.~Caianiello' dell'Universit\`{a} and Gruppo Collegato INFN, Salerno, Italy\\
$^{29}$ Dipartimento DISAT del Politecnico and Sezione INFN, Turin, Italy\\
$^{30}$ Dipartimento di Scienze MIFT, Universit\`{a} di Messina, Messina, Italy\\
$^{31}$ Dipartimento Interateneo di Fisica `M.~Merlin' and Sezione INFN, Bari, Italy\\
$^{32}$ European Organization for Nuclear Research (CERN), Geneva, Switzerland\\
$^{33}$ Faculty of Electrical Engineering, Mechanical Engineering and Naval Architecture, University of Split, Split, Croatia\\
$^{34}$ Faculty of Engineering and Science, Western Norway University of Applied Sciences, Bergen, Norway\\
$^{35}$ Faculty of Nuclear Sciences and Physical Engineering, Czech Technical University in Prague, Prague, Czech Republic\\
$^{36}$ Faculty of Physics, Sofia University, Sofia, Bulgaria\\
$^{37}$ Faculty of Science, P.J.~\v{S}af\'{a}rik University, Ko\v{s}ice, Slovak Republic\\
$^{38}$ Frankfurt Institute for Advanced Studies, Johann Wolfgang Goethe-Universit\"{a}t Frankfurt, Frankfurt, Germany\\
$^{39}$ Fudan University, Shanghai, China\\
$^{40}$ Gangneung-Wonju National University, Gangneung, Republic of Korea\\
$^{41}$ Gauhati University, Department of Physics, Guwahati, India\\
$^{42}$ Helmholtz-Institut f\"{u}r Strahlen- und Kernphysik, Rheinische Friedrich-Wilhelms-Universit\"{a}t Bonn, Bonn, Germany\\
$^{43}$ Helsinki Institute of Physics (HIP), Helsinki, Finland\\
$^{44}$ High Energy Physics Group,  Universidad Aut\'{o}noma de Puebla, Puebla, Mexico\\
$^{45}$ Horia Hulubei National Institute of Physics and Nuclear Engineering, Bucharest, Romania\\
$^{46}$ HUN-REN Wigner Research Centre for Physics, Budapest, Hungary\\
$^{47}$ Indian Institute of Technology Bombay (IIT), Mumbai, India\\
$^{48}$ Indian Institute of Technology Indore, Indore, India\\
$^{49}$ INFN, Laboratori Nazionali di Frascati, Frascati, Italy\\
$^{50}$ INFN, Sezione di Bari, Bari, Italy\\
$^{51}$ INFN, Sezione di Bologna, Bologna, Italy\\
$^{52}$ INFN, Sezione di Cagliari, Cagliari, Italy\\
$^{53}$ INFN, Sezione di Catania, Catania, Italy\\
$^{54}$ INFN, Sezione di Padova, Padova, Italy\\
$^{55}$ INFN, Sezione di Pavia, Pavia, Italy\\
$^{56}$ INFN, Sezione di Torino, Turin, Italy\\
$^{57}$ INFN, Sezione di Trieste, Trieste, Italy\\
$^{58}$ Inha University, Incheon, Republic of Korea\\
$^{59}$ Institute for Gravitational and Subatomic Physics (GRASP), Utrecht University/Nikhef, Utrecht, Netherlands\\
$^{60}$ Institute of Experimental Physics, Slovak Academy of Sciences, Ko\v{s}ice, Slovak Republic\\
$^{61}$ Institute of Physics, Homi Bhabha National Institute, Bhubaneswar, India\\
$^{62}$ Institute of Physics of the Czech Academy of Sciences, Prague, Czech Republic\\
$^{63}$ Institute of Space Science (ISS), Bucharest, Romania\\
$^{64}$ Institut f\"{u}r Kernphysik, Johann Wolfgang Goethe-Universit\"{a}t Frankfurt, Frankfurt, Germany\\
$^{65}$ Instituto de Ciencias Nucleares, Universidad Nacional Aut\'{o}noma de M\'{e}xico, Mexico City, Mexico\\
$^{66}$ Instituto de F\'{i}sica, Universidade Federal do Rio Grande do Sul (UFRGS), Porto Alegre, Brazil\\
$^{67}$ Instituto de F\'{\i}sica, Universidad Nacional Aut\'{o}noma de M\'{e}xico, Mexico City, Mexico\\
$^{68}$ iThemba LABS, National Research Foundation, Somerset West, South Africa\\
$^{69}$ Jeonbuk National University, Jeonju, Republic of Korea\\
$^{70}$ Johann-Wolfgang-Goethe Universit\"{a}t Frankfurt Institut f\"{u}r Informatik, Fachbereich Informatik und Mathematik, Frankfurt, Germany\\
$^{71}$ Korea Institute of Science and Technology Information, Daejeon, Republic of Korea\\
$^{72}$ KTO Karatay University, Konya, Turkey\\
$^{73}$ Laboratoire de Physique Subatomique et de Cosmologie, Universit\'{e} Grenoble-Alpes, CNRS-IN2P3, Grenoble, France\\
$^{74}$ Lawrence Berkeley National Laboratory, Berkeley, California, United States\\
$^{75}$ Lund University Department of Physics, Division of Particle Physics, Lund, Sweden\\
$^{76}$ Nagasaki Institute of Applied Science, Nagasaki, Japan\\
$^{77}$ Nara Women{'}s University (NWU), Nara, Japan\\
$^{78}$ National and Kapodistrian University of Athens, School of Science, Department of Physics , Athens, Greece\\
$^{79}$ National Centre for Nuclear Research, Warsaw, Poland\\
$^{80}$ National Institute of Science Education and Research, Homi Bhabha National Institute, Jatni, India\\
$^{81}$ National Nuclear Research Center, Baku, Azerbaijan\\
$^{82}$ National Research and Innovation Agency - BRIN, Jakarta, Indonesia\\
$^{83}$ Niels Bohr Institute, University of Copenhagen, Copenhagen, Denmark\\
$^{84}$ Nikhef, National institute for subatomic physics, Amsterdam, Netherlands\\
$^{85}$ Nuclear Physics Group, STFC Daresbury Laboratory, Daresbury, United Kingdom\\
$^{86}$ Nuclear Physics Institute of the Czech Academy of Sciences, Husinec-\v{R}e\v{z}, Czech Republic\\
$^{87}$ Oak Ridge National Laboratory, Oak Ridge, Tennessee, United States\\
$^{88}$ Ohio State University, Columbus, Ohio, United States\\
$^{89}$ Physics department, Faculty of science, University of Zagreb, Zagreb, Croatia\\
$^{90}$ Physics Department, Panjab University, Chandigarh, India\\
$^{91}$ Physics Department, University of Jammu, Jammu, India\\
$^{92}$ Physics Program and International Institute for Sustainability with Knotted Chiral Meta Matter (SKCM2), Hiroshima University, Hiroshima, Japan\\
$^{93}$ Physikalisches Institut, Eberhard-Karls-Universit\"{a}t T\"{u}bingen, T\"{u}bingen, Germany\\
$^{94}$ Physikalisches Institut, Ruprecht-Karls-Universit\"{a}t Heidelberg, Heidelberg, Germany\\
$^{95}$ Physik Department, Technische Universit\"{a}t M\"{u}nchen, Munich, Germany\\
$^{96}$ Politecnico di Bari and Sezione INFN, Bari, Italy\\
$^{97}$ Research Division and ExtreMe Matter Institute EMMI, GSI Helmholtzzentrum f\"ur Schwerionenforschung GmbH, Darmstadt, Germany\\
$^{98}$ Saga University, Saga, Japan\\
$^{99}$ Saha Institute of Nuclear Physics, Homi Bhabha National Institute, Kolkata, India\\
$^{100}$ School of Physics and Astronomy, University of Birmingham, Birmingham, United Kingdom\\
$^{101}$ Secci\'{o}n F\'{\i}sica, Departamento de Ciencias, Pontificia Universidad Cat\'{o}lica del Per\'{u}, Lima, Peru\\
$^{102}$ Stefan Meyer Institut f\"{u}r Subatomare Physik (SMI), Vienna, Austria\\
$^{103}$ SUBATECH, IMT Atlantique, Nantes Universit\'{e}, CNRS-IN2P3, Nantes, France\\
$^{104}$ Sungkyunkwan University, Suwon City, Republic of Korea\\
$^{105}$ Suranaree University of Technology, Nakhon Ratchasima, Thailand\\
$^{106}$ Technical University of Ko\v{s}ice, Ko\v{s}ice, Slovak Republic\\
$^{107}$ The Henryk Niewodniczanski Institute of Nuclear Physics, Polish Academy of Sciences, Cracow, Poland\\
$^{108}$ The University of Texas at Austin, Austin, Texas, United States\\
$^{109}$ Universidad Aut\'{o}noma de Sinaloa, Culiac\'{a}n, Mexico\\
$^{110}$ Universidade de S\~{a}o Paulo (USP), S\~{a}o Paulo, Brazil\\
$^{111}$ Universidade Estadual de Campinas (UNICAMP), Campinas, Brazil\\
$^{112}$ Universidade Federal do ABC, Santo Andre, Brazil\\
$^{113}$ Universitatea Nationala de Stiinta si Tehnologie Politehnica Bucuresti, Bucharest, Romania\\
$^{114}$ University of Cape Town, Cape Town, South Africa\\
$^{115}$ University of Derby, Derby, United Kingdom\\
$^{116}$ University of Houston, Houston, Texas, United States\\
$^{117}$ University of Jyv\"{a}skyl\"{a}, Jyv\"{a}skyl\"{a}, Finland\\
$^{118}$ University of Kansas, Lawrence, Kansas, United States\\
$^{119}$ University of Liverpool, Liverpool, United Kingdom\\
$^{120}$ University of Science and Technology of China, Hefei, China\\
$^{121}$ University of South-Eastern Norway, Kongsberg, Norway\\
$^{122}$ University of Tennessee, Knoxville, Tennessee, United States\\
$^{123}$ University of the Witwatersrand, Johannesburg, South Africa\\
$^{124}$ University of Tokyo, Tokyo, Japan\\
$^{125}$ University of Tsukuba, Tsukuba, Japan\\
$^{126}$ Universit\"{a}t M\"{u}nster, Institut f\"{u}r Kernphysik, M\"{u}nster, Germany\\
$^{127}$ Universit\'{e} Clermont Auvergne, CNRS/IN2P3, LPC, Clermont-Ferrand, France\\
$^{128}$ Universit\'{e} de Lyon, CNRS/IN2P3, Institut de Physique des 2 Infinis de Lyon, Lyon, France\\
$^{129}$ Universit\'{e} de Strasbourg, CNRS, IPHC UMR 7178, F-67000 Strasbourg, France, Strasbourg, France\\
$^{130}$ Universit\'{e} Paris-Saclay, Centre d'Etudes de Saclay (CEA), IRFU, D\'{e}partment de Physique Nucl\'{e}aire (DPhN), Saclay, France\\
$^{131}$ Universit\'{e}  Paris-Saclay, CNRS/IN2P3, IJCLab, Orsay, France\\
$^{132}$ Universit\`{a} degli Studi di Foggia, Foggia, Italy\\
$^{133}$ Universit\`{a} del Piemonte Orientale, Vercelli, Italy\\
$^{134}$ Universit\`{a} di Brescia, Brescia, Italy\\
$^{135}$ Variable Energy Cyclotron Centre, Homi Bhabha National Institute, Kolkata, India\\
$^{136}$ Warsaw University of Technology, Warsaw, Poland\\
$^{137}$ Wayne State University, Detroit, Michigan, United States\\
$^{138}$ Yale University, New Haven, Connecticut, United States\\
$^{139}$ Yonsei University, Seoul, Republic of Korea\\
$^{140}$  Zentrum  f\"{u}r Technologie und Transfer (ZTT), Worms, Germany\\
$^{141}$ Affiliated with an institute covered by a cooperation agreement with CERN\\
$^{142}$ Affiliated with an international laboratory covered by a cooperation agreement with CERN.\\

\end{flushleft} 

\end{document}